\documentclass[aps,showpacs,twocolumn,footinbib,%
superscriptaddress]{revtex4-1}
\usepackage{amsmath,amssymb,bm,graphicx,hyperref}
\usepackage{latexsym}
\usepackage{epsfig}
\usepackage{amsfonts}
\usepackage{bm}
\usepackage{braket}
\usepackage{ascmac}
\usepackage{color}
\usepackage{pstricks}%
\usepackage{pst-eps}%
\usepackage{pst-plot}%
\usepackage{pst-node}%
\usepackage{pst-poly}

\renewcommand{\i}{{\rm i}}
\newcommand{\e}{{\rm e}}

\renewcommand{\d}{{\rm d}}

\begin{document}
\title{
Extraction of topological information in Tomonaga-Luttinger liquids}
\author{Masaaki Nakamura }
\affiliation{
Department of Physics, Ehime University Bunkyo-cho 2-5, Matsuyama, Ehime
790-8577, Japan}
\author{Shunsuke C. Furuya}
\affiliation{
Condensed Matter Theory Laboratory, RIKEN, Wako, Saitama 351-0198, Japan}
\date{\today}

\begin{abstract}
We discuss expectation values of the twist operator $U$ appearing in the
Lieb-Schultz-Mattis theorem (or the polarization operator for periodic
systems) in excited states of the one-dimensional correlated systems
$z_L^{(q,\pm)}\equiv\braket{\Psi_{q/2}^{\pm}|U^q|\Psi_{q/2}^{\pm}}$,
where $\ket{\Psi_{p}^{\pm}}$ denotes the excited states given by linear
combinations of momentum $2pk_{\rm F}$ with parity $\pm 1$.  We found
that $z_L^{(q,\pm)}$ gives universal values $\pm 1/2$ on the
Tomonaga-Luttinger (TL) fixed point, and its signs identify the topology
of the dominant phases.  Therefore, this expectation value changes
between $\pm 1/2$ discontinuously at a phase transition point with the
U(1) or SU(2) symmetric Gaussian universality class.  This means that
$z_L^{(q,\pm)}$ extracts the topological information of TL liquids.  We
explain these results based on the free-fermion picture and the
bosonization theory, and also demonstrate them in several physical
systems.
\end{abstract}

\maketitle 

\section{Introduction}
In many body quantum systems, it is important to investigate structures
of low-energy spectra such as the existence of energy gaps and the
degeneracy of ground states. These structures of energy spectra
characterize the physical properties of the systems such as metals or
insulators, and dominant phases.

The Lieb-Schultz-Mattis (LSM) theorem plays an important role in the
study of such properties in one-dimensional (1D) lattice systems
\cite{Lieb-S-M,Affleck-L,Affleck,Oshikawa-Y-A,Yamanaka-O-A}. In the LSM
theorem, the possibility of opening an energy gap in a parity and
translationally symmetric system is related to the orthogonality of a
non-degenerate ground state in a finite-size system $\ket{\Psi_0}$ and a
variational excited state $U^q\ket{\Psi_0}$.  Here, $U$ is the twist
operator which creates the $O(1/L)$ excitation in a finite $L$ size
system. For fermion systems, that is defined by
\begin{equation}
 U=\exp\biggl(\frac{2\pi{\rm i}}{L}\sum_{j=1}^L j n_j\biggr),
\end{equation}
where $n_j$ is the density operator at site $j$. For spin systems, the
twist operator is defined by replacing the density operator $n_j$ by the
spin operator $S_j^z$.  It is well known that as a generalization of the
original LSM theorem ($q=1$), the necessary condition for the appearance
of gapped states with $q$-fold degenerate ground states is given by
$q(S-m)=$ integer where $S$ and $m$ are the spin and the magnetization
per unit cell \cite{Oshikawa-Y-A}. In this way
\begin{equation}
 z_L^{(q)}=\braket{\Psi_0|U^q|\Psi_0}\label{z0def}
\end{equation}
is the essential index in the LSM theorem.

On the other hand, the same quantity $z_L^{(q)}$ is also introduced by
Resta from an argument of electric polarization. He introduced
$z_L^{(1)}$ to define the expectation value of the center-of-mass
operator $\frac{1}{L}\sum_{j=1}^L j n_j$ in periodic systems
\cite{Resta,Resta-S1999,Resta2000}. This notion was also extended to
$q$-fold degenerate systems \cite{Aligia-O}. It is well known that an
insulator is distinguished from a conductor at zero temperature by its
vanishing dc conductivity (Drude weight)\cite{Kohn},
whereas, $z_L^{(q)}$ distinguishes not only metals and insulators, but
also ``topology'' of insulators by its sign, such as band or Mott
insulators. Thus $z_L^{(q)}$ plays the role of order parameters and also
probes to detect topological phase
transitions \cite{Nakamura-V,Nakamura-T}.

In this paper, we turn our attention to the following expectation value
of $U$:
\begin{equation}
 z_L^{(q,\pm)}=\braket{\Psi_{q/2}^{\pm}|U^q|\Psi_{q/2}^{\pm}},
  \label{zdef}
\end{equation}
where $\ket{\Psi_p^{\pm}}$ denotes linear combinations of excited states
with momenta $2pk_{\rm F}$ and $-2pk_{\rm F}$, and with parity
$\mathcal{P}\ket{\Psi_{q/2}^{\pm}}=\pm\ket{\Psi_{q/2}^{\pm}}$.  Here
$k_{\rm F}$ is the Fermi momentum with $qk_{\rm F}=n\pi$ ($n$: integer).
This is as an extension of Eq.~(\ref{z0def}), but, as will be shown
later, it extracts the topological information of 1D quantum systems at
the limit of the Tomonaga-Luttinger (TL) fixed point as the universal
values $z_L^{(q,\pm)}=\pm 1/2$, whereas $z_L^{(q)}$ becomes zero. This
is essentially different from the property of $z_L^{(q)}$ whose sign is
determined in the gapped fixed points.

This paper is organized as follows. In Sec.~\ref{sec:free_fermion}, we
discuss the properties of $z_L^{(q,\pm)}$ in the free fermions.  In
Sec.~\ref{sec:bosonization}, we discuss the interacting systems based on
the TL model and bosonization of the twist operator.  In
Sec.~\ref{sec:models}, we demonstrate the properties in several physical
systems based on the exact diagonalization (ED). Finally summary and
discussions are given in Sec.~\ref{sec:summary}. Throughout this paper,
the lattice constant and the Planck constant are set to be unity.

\section{Free fermion picture}
\label{sec:free_fermion}
\begin{figure}[t]
\includegraphics[width=9cm]{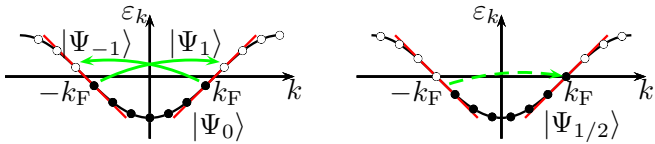}
\caption{Dispersion relations of the ground state $\ket{\Psi_0}$ and the
 excited states $\ket{\Psi_p}$ in finite-size systems. States with
 half-odd integers $p$ are realized in the antiperiodic boundary
 condition, where the wave number $k=\frac{2\pi}{L}m$ with a half-odd
 integer $m$.  In particular, excited states $\ket{\Psi_{\pm 1/2}}$ are
 realized as doubly-degenerate ground states in the antiperiodic
 boundary condition.  The momentum transfer of $\ket{\Psi_{1/2}}$ is
 regarded as $2k_{\rm F}$ by $1/2$ fermion.}
\label{fig:disp}
\end{figure}

First, we consider the properties of Eq.~(\ref{zdef}) in free-fermion
systems.  It follows from the relation of the creation operators in the
real and the momentum spaces, and the twist operator
\begin{equation}
 Uc_j^{\dag}U^{-1}=c_j^{\dag}\e^{\i\frac{2\pi}{L}j},\quad
 Uc_k^{\dag}U^{-1}=c_{k+2\pi/L}^{\dag},
\end{equation}
that $U$ creates the momentum shift $\frac{2\pi}{L}$.  This means that
it creates momentum transfer $2k_{\rm F}$ with respect to the ground
state where the fermion states between $k=-k_{\rm F}$ and $k=k_{\rm F}$
are occupied.  Now we introduce the following excited states:
\begin{equation}
 \ket{\Psi_p^{\pm}}
 \equiv\frac{1}{\sqrt{2}}(\ket{\Psi_{+p}}\pm\ket{\Psi_{-p}}),
 \label{excited_p}
\end{equation}
where $\ket{\Psi_{p}}$ is a state with a momentum $2pk_{\rm F}$ (see
Fig.~\ref{fig:disp}).  $\ket{\Psi_p^{\pm}}$ are eigenstates of the
parity operation
$\mathcal{P}\ket{\Psi_{p}^{\pm}}=\pm\ket{\Psi_{p}^{\pm}}$, since
$\mathcal{P}\ket{\Psi_{p}}=\ket{\Psi_{-p}}$.  The momentum is restricted
by a condition $2pk_{\rm F}=n\pi$ ($n$: integer), since the parity
operation $\mathcal{P}$ ($\mathcal{P}c_i\mathcal{P}^{-1}=c_{L+1-i}$)
commutes with the one-site shift operation $\mathcal{T}$
($\mathcal{T}c_i\mathcal{T}^{-1}=c_{i+1}$) only when the eigenvalue of
$\mathcal{T}$ is a real number $\pm 1$. In this situation, the
expectation value of the twist operator with an integer $q$ becomes
\begin{align}
 \bra{\Psi_{p}^\pm}U^q\ket{\Psi_{p}^\pm}
 =&\frac{1}{2}\braket{\Psi_0|
 (U^{q}\pm U^{q-2p}\pm U^{q+2p}+U^{q})|\Psi_0}
 \nonumber
 \\
 =&\pm\frac{1}{2}\delta_{q,2p},\quad (L\to\infty).
 \label{z03}
\end{align}
Here terms with the finite power of $U$ vanish in the $L\to\infty$ limit
due to the LSM theorem which insists that $\ket{\Psi_0}$ and
$U^{q}\ket{\Psi_0}$ are orthogonal in the gapless state.  Thus it turns
out that $z_L^{(q,\pm)}=\pm 1/2$ and the signs identify parities of the
wave function of the excited states (\ref{excited_p}) with $p=q/2$. Note
that states with half-odd integers $p$ are realized in antiperiodic
boundary conditions, since the wave numbers are given by
$k=\frac{2\pi}{L}m$ with half-odd integers $m$ as shown in
Fig.~\ref{fig:disp}.

\section{Bosonization}\label{sec:bosonization}
Next, we demonstrate that the above property of $z_L^{(q,\pm )}$ is
unchanged in interacting cases.  To this end we consider the
bosonization of the twist operator.  In 1D systems, the low-energy
excitations are described as TL
liquids \cite{Haldane1981,Giamarchi,Delft-S}. The Hamiltonian of the
interacting fermions is given by the Gaussian model,
\begin{equation}
 \mathcal{H}_{\rm TL} = v \int \d x 
  \left[\frac{\pi K}{2} \Pi^2  + \frac{1}{2\pi K }
   \left(\frac{\partial \phi}{\partial x}\right)^2\right],
  \label{Gaussian}
\end{equation}
where $v$ and $K$ are the sound velocity and the TL parameter,
respectively. The phase fields satisfy $\left[ \phi (x), \Pi
(y)\right]=\i\delta (x-y)$ and their mode expansions of the phase fields
are given by
\begin{align}
\phi (x)
 =&\frac{\i\pi }{L} \sum_{k\neq 0} \frac{1}{k} \e^{-\alpha
 \frac{|k|}{2}-\i kx} \left[\rho_+ (k) + \rho_- (k)\right]
  + \frac{N\pi x}{L}+Q,\\
 \Pi (x)
 =& -\frac{1}{L}\sum_{k\neq 0} \e^{-\alpha \frac{|k|}{2}-\i kx}
 \left[\rho_+ (k) - \rho_- (k)\right] -\frac{M}{L},
\end{align}
with the cutoff $\alpha$. The density operators satisfy the following
commutation relation:
\begin{equation}
 \left[\rho_r (-k), \rho_{r'} (k')\right] =  r\frac{kL}{2\pi}
  \delta_{rr'}\delta_{kk'} \quad (r,r'=+,-),
\end{equation}
and $\rho_+(-k)\ket{\Psi_0}=\rho_-(k)\ket{\Psi_0}=0$ for $k>0$.  The
zero mode satisfies the relation $[Q,M]=-\i$.  The effects of the
interactions are renormalized into the TL parameter $K$, whereas $K=1$
is for the free fermions.  Usually, the low-energy Hamiltonian also
includes a non-linear term as $\mathcal{H}=\mathcal{H}_{\rm TL}+\frac{2
g}{(2\pi\alpha)^2} \int_0^L {\rm d}x \cos[2q\phi(x)]$ which opens an
energy gap when it is relevant in the renormalization group
analysis. Therefore Hamiltonian (\ref{Gaussian}) is realized just on the
transition point with the Gaussian universality class ($g=0$).

The center-of-mass operator is bosonized using the partial integration
as
\begin{align}
\frac{2\pi}{L}\sum_{j=1}^L j n_j
\to &\frac{2\pi}{L}\int_0^L\d x x\frac{1}{\pi}\partial_x\phi(x)\\
&=2\phi(L)-N\pi-2Q,
\end{align}
where we have ignored the $2k_{\rm F}$-umklapp term.  Then the normal
ordered representation becomes\cite{Delft-S} (see Appendix
\ref{sec:deriv_boson_U})
\begin{align}
 U^q\to&\mathcal{U}(q,K)\equiv 
 \exp\left[\i q(2\phi(L)-N\pi-2Q)\right]\label{bosonized_U0}\\
 \simeq&:\exp\left[\i 2q \phi(L)\right]:
 \left(\frac{2\pi\alpha}{L}\right)^{q^2K}.
\label{bosonized_U}
\end{align}
If we set the cut-off parameter $\alpha$ to the order of the lattice
constant $\sim 1$, this result describes the $O(1/L)$ excitation in the
LSM theorem. This is also consistent with the conjecture
$z_L^{(q)}\propto\braket{\cos 2q\phi}$ discussed in
Ref.~\onlinecite{Nakamura-V}, since $z_L^{(q)}$ is a real number under
the parity symmetry $\phi\to-\phi$. We can also confirm that the
bosonized representation satisfies the relation (see Appendix
\ref{sec:deriv_boson_U})
\begin{equation}
 \mathcal{U}(q,K)\mathcal{U}(p,K)=\mathcal{U}(p+q,K).
  \label{UUU}
\end{equation}
For the interacting case with the forward scattering $K\neq 1$, it
follows from the concept of TL liquids that the state with $2qk_{\rm F}$
momentum $\ket{\Psi_q}$ is given by
\begin{equation}
 \ket{\Psi_q(K)}=
 \mathcal{U}(q,K)\ket{\Psi_0(K)},
 \label{Psi_q_K}
\end{equation}
where $\ket{\Psi_0(K)}$ is the ground state.  Therefore, the relation
for free fermions (\ref{z03}) is also applicable to the interacting case
(\ref{Gaussian}) only by a replacement $U^q\to\mathcal{U}(q,K)$. Thus
within the low-energy approximation, the values $z_L^{(q,\pm)}=\pm 1/2$
turn out to be universal for the TL liquids with $K\neq 1$. This result
does not depend on the detailed form of $\mathcal{U}(q,K)$ as long as
the relation (\ref{UUU}) is satisfied.

In the conformal field theory (CFT), expectation values of one-point
operators in finite-size systems are evaluated
as \cite{Cardy86,Reinicke}, (see Appendix~\ref{sec:three_point})
\begin{align}
 \langle {\cal O}_i|{\cal O}_j(\sigma)|{\cal O}_i\rangle
  =&
  C_{iji}
  \left(\frac{2\pi}{L}\right)^{x_j},\label{correction_to_FSS}
\end{align}
where $x_j$ is the scaling dimension of the $j$th operator, and
$C_{iji}$ is the operator product expansion (OPE) coefficient defined
as,
\begin{equation}
 {\cal O}_i(\sigma_1)
  {\cal O}_j(\sigma_2)
  =\sum_k\frac{C_{ijk}}{|\sigma_1-\sigma_2|^{x_i+x_j-x_k}}
  {\cal O}_k(\sigma_2).
\end{equation}
In the present case, the excited states $\ket{\Psi_{q/2}^{\pm}}$ are
eigenstates of
${\cal O}_1(\sigma)\equiv:\!\cos[q\phi(\sigma)]\!:$ and
${\cal O}_2(\sigma)\equiv:\!\sin[q\phi(\sigma)]\!:$,
respectively. In addition,
${\cal O}_3(\sigma)\equiv:\!\cos[2q\phi(\sigma)]\!:$ is related to the
twist operator as ${\cal O}_3(L)\propto\mathcal{U}(q,K)$. The scaling
dimensions are $x_1=x_2=q^2K/4$ and $x_3=q^2K$. The OPE coefficients are
$C_{131}=+1/2$ and $C_{232}=-1/2$ (see Appendix~\ref{sec:OPE}). Then,
the formula, Eq.~(\ref{correction_to_FSS}), seems to explain
Eq.~(\ref{z03}), but the size dependence $(2\pi/L)^{x_3}$ remains. This
discrepancy is because the bosonized operator ${\cal O}_3(L)$ is no
longer a local field, so that Eq.~(\ref{correction_to_FSS}) is not
applicable to the present case.

\begin{figure}[t]
\begin{center}
\includegraphics[width=7.5cm]{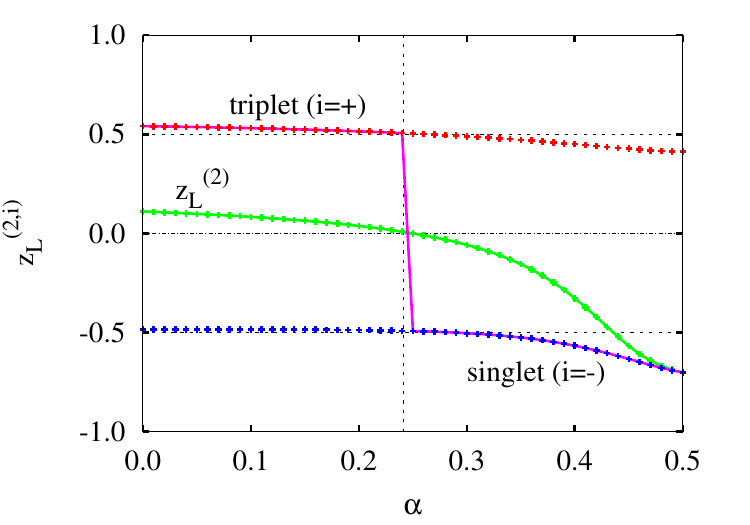}
\end{center}
\caption{$z_L^{(2,\pm)}$ of the $S=1/2$ $J_1$-$J_2$ spin chain for the
$L=28$ system obtained by ED (red and blue lines).  If we calculate the
first excited state without classifying the Hilbert space by parity, the
value change discontinuously between $\pm1/2$ at the gapless-dimer
transition point $\alpha_{\rm c}=0.2411$ (magenta line).  On the other
hand, $z_L^{(2)}$ changes continuously and becomes zero at $\alpha_{\rm
c}$ (green line). $z_L^{(2,\pm)}$ converges to $\pm 1/2$ for the gapless
region, while to a finite value for the dimer
region.}\label{fig:z1_J1J2XXZ}
\end{figure}

\begin{figure}[t]
\begin{center}
\includegraphics[width=7.5cm]{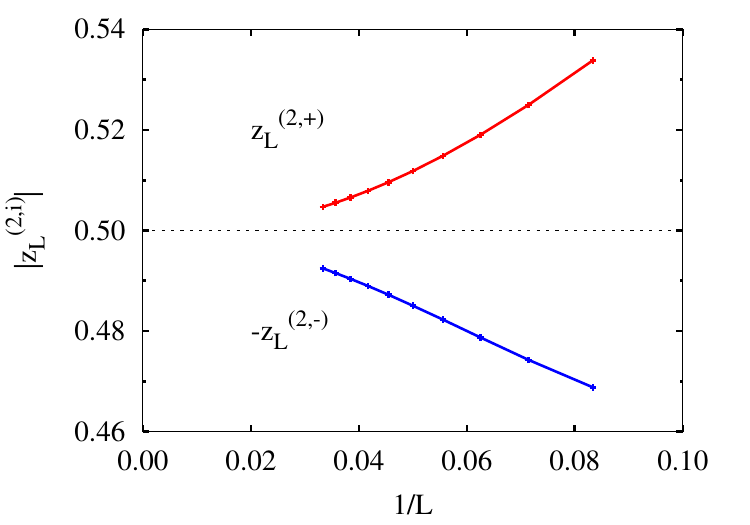}
\end{center}
\caption{System size dependence ($L=12$-$30$) of $z_L^{(2,\pm)}$ of the
$S=1/2$ $J_1$-$J_2$ spin chain at the gapless-dimer transition point
$\alpha_{\rm c}=0.2411$. This shows that $z_L^{(2,\pm)}$ has the size
dependence $O(1/L)$, and approach to $\pm 1/2$ in the $L\to\infty$
limit.}
\label{fig:size_z1_J1J2XXZ}
\end{figure}


\section{Physical Systems}
\label{sec:models}
\subsection{The $S=1/2$ $J_1$-$J_2$ spin chain}
In the rest of this paper, we demonstrate the above argument in several
models based on ED.  As the first example, we consider the $S=1/2$
antiferromagnetic Heisenberg chain with the next-nearest-neighbor
exchanges,
\begin{equation}
 \mathcal{H}
 =\sum_{i=1}^L
 \left[\bm{S}_i\cdot\bm{S}_{i+1}
 +\alpha\bm{S}_i\cdot\bm{S}_{i+2}\right].
 \label{J1J2}
\end{equation}
In this system, a phase transition between the gapless state and the
dimer state occurs at $\alpha_{\rm c}=0.2411$ \cite{Okamoto-N}. This
critical point belongs to the universality class of the SU(2) symmetric
Gaussian model, and is identified by the level-crossing of the
singlet-triplet excitation energies. These excited states correspond to
$\ket{\Psi_{1}^{\pm}}$.  The critical point can also be determined by
the condition $z_L^{(2)}=0$ \cite{Nakamura-V,Nakamura-T} which is
obtained by the ground-state expectation value of $U^2$ by ED for the
$L=28$ system, as shown in Fig.~\ref{fig:z1_J1J2XXZ}.

Now we turn our attention to $z_L^{(2,\pm)}$ for the singlet
$\ket{\Psi_{1}^{+}}$ and the triplet $\ket{\Psi_{1}^{-}}$ states
corresponding to the dimer and the gapless states, respectively.
According to the results of ED in Figs.~\ref{fig:z1_J1J2XXZ} and
\ref{fig:size_z1_J1J2XXZ}, $z_L^{(2,\pm)}=\pm 1/2$ at the critical point
$\alpha=\alpha_{\rm c}$ with the size dependence $O(1/L)$.  If we
calculate the first excited state without classifying the Hilbert space
by parity and/or spin-reversal symmetries, the the expectation value
changes discontinuously at $\alpha_{\rm c}$.  For the gapless region
$\alpha<\alpha_{\rm c}$, the values $z_L^{(2,\pm)}=\pm 1/2$ are almost
constant, while they deviate from $\pm 1/2$ for the dimer regions
$\alpha>\alpha_{\rm c}$. This indicates that for the gapped region
$\braket{\Psi_0|U^q|\Psi_0}\neq 0$ due to the LSM theorem, so that
Eq.~(\ref{z03}) is not satisfied. For the Majumder-Gorsh point
$\alpha=0.5$ where the system is fully dimerized
\cite{Majumder-G,Majumder,AKLT1,AKLT2}, the expectation values of $U^2$
with respect to the two states $\ket{\Psi_0}$ and $\ket{\Psi_1^-}$ give
the same value $z_L^{(2,-)}=z_L^{(2)}\simeq-[\cos(2\pi/L)]^{L/2}$.

\subsection{The $S=1$ spin chain}
The next example is the $S=1$ Heisenberg chain with the single-ion
anisotropy,
\begin{align}
 \mathcal{H}
 =\sum_{i=1}^L
 \left[\bm{S}_i\cdot\bm{S}_{i+1}+D(S_i^z)^2\right].
 \label{S1D}
\end{align}
This model undergoes a U(1) Gaussian-type phase transition from the
Haldane-gap state \cite{Haldane1983a,Haldane1983b} to the large-$D$ (or
trivial) state at $D_{\rm c}=0.968\pm 0.001$
\cite{Chen-H-S2000,Chen-H-S2003,Chen-H-S2008}. This transition point is
determined by the level-crossing of low-energy spectra of
$\ket{\Psi_{1/2}^{\pm}}$ obtained with antiperiodic boundary conditions
\cite{Kitazawa}. The twisted boundary conditions play a role to make
artificial low-energy excitations that degenerate with the Haldane
$\ket{\Psi_{1/2}^{-}}$ and large-$D$ $\ket{\Psi_{1/2}^{+}}$ ground
states, respectively. The transition point $D_{\rm c}$ can also be
identified by $z_L^{(1)}=0$ \cite{Nakamura-B-S}.

\begin{figure}[t]
\includegraphics[width=7.5cm]{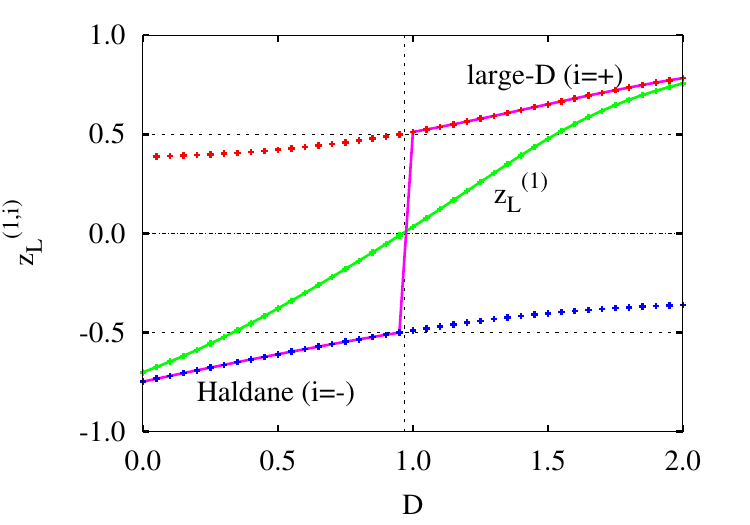}
\caption{$z_L^{(1,\pm)}$ of the $S=1$ XXZ chain for the $L=18$ system
obtained by ED (red and blue lines).  If we calculate the first excited
state without classifying the Hilbert space by parity, the values change
discontinuously between $\pm1/2$ at the Haldane-large-$D$ transition
point $D_{\rm c}=0.968$ (magenta line).  On the other hand, $z_L^{(1)}$
changes continuously and becomes zero at $D_{\rm c}$ (green line).
$z_L^{(1,\pm)}$ do not converge to $\pm 1/2$ for the gapped regions
$D\neq D_{\rm c}$.
}\label{fig:z1_S1XXZD}
\end{figure}

\begin{figure}[t]
\begin{center}
\includegraphics[width=7.5cm]{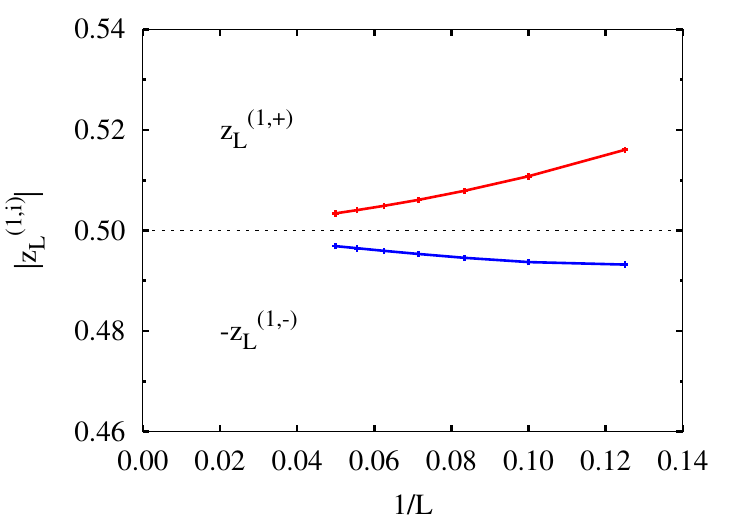}

\end{center}
\caption{System size dependence ($L=8$-$20$) of $z_L^{(1,\pm)}$ of the
$S=1$ XXZ chain at the Haldane-large-$D$ transition point $D_{\rm
c}=0.968$.  This shows that $z_L^{(1,\pm)}$ has the size dependence
$O(1/L)$, and approach to $\pm 1/2$ in the $L\to\infty$
limit.}\label{fig:size_z1_S1XXZD}
\end{figure}

It follows from the results obtained by ED shown in
Figs.~\ref{fig:z1_S1XXZD} and \ref{fig:size_z1_S1XXZD},
$z_L^{(1,\pm)}\pm 1/2$ with the size dependence $O(1/L)$.  The excited
states correspond to the Haldane ($\ket{\Psi_{1/2}^{-}}$) and the
large-$D$ ($\ket{\Psi_{1/2}^{+}}$) phases, respectively.  Unlike the
case of the $S=1/2$ $J_1$-$J_2$ spin chain, $z_L^{(1,\pm)}$ deviates
from $\pm1/2$ away from $D_{\rm c}$, because both two regions $D\gtrless
D_{\rm c}$ are gapped states.

\subsection{The extended Hubbard model}
As an electron system, we consider the 1D extended Hubbard model at
half-filling and zero magnetic field,
\begin{align}
{\cal H}=
\sum_{i=1}^L
 \biggl[-t\sum_{s=\uparrow,\downarrow}
 (c^{\dag}_{is} c_{i+1,s}&+\mbox{H.c.})\nonumber\\
+&U n_{i\uparrow}n_{i\downarrow}
 +V n_{i}n_{i+1} \biggr],
 \label{eqn:tUV}
\end{align}
where $c_{is}$ ($c_{is}^\dag$) is the electron annihilation (creation)
operator for spin $s=\uparrow,\downarrow$. The number operators are
defined by $n_{is}\equiv c^{\dag}_{is} c_{is}^{\mathstrut}$ and
$n_i\equiv n_{i\uparrow}+n_{i\downarrow}$.  According to the analysis of
the excitation spectra \cite{Nakamura1999,Nakamura2000}, the U(1)
Gaussian transition in the charge part, and the SU(2) symmetric spin-gap
transition take place independently near the $U=2V$ line with
$0<U<U_{\rm c}$, where $U_{\rm c}$ is the tricritical point.  Therefore,
there appear three phases around $U=2V$. Those are spin-density-wave
(SDW), bond-charge-density-wave (BCDW), and charge-density-wave (CDW)
phases.

\begin{figure}[t]
\begin{center}
\includegraphics[width=7.5cm]{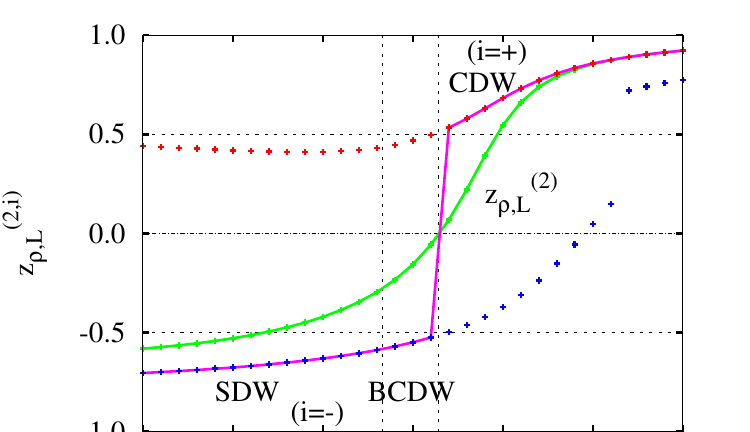}\\
\includegraphics[width=7.5cm]{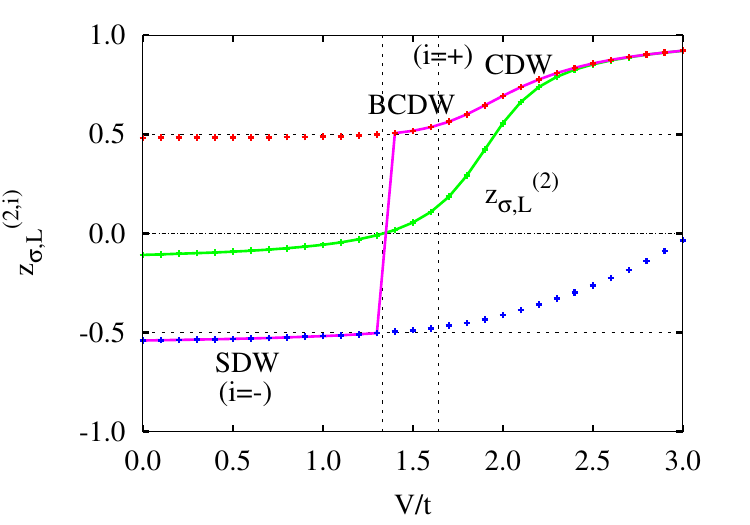}
\end{center}
\caption{$z_{\nu,L}^{(2,\pm)}$ of the extended Hubbard model for the
 charge ($\nu=\rho$) and the spin ($\nu=\sigma$) sectors for the $L=14$
 system at $U/t=3$ obtained by ED (red and blue lines).
 At the BCDW-CDW (SDW-BCDW) boundary, we get $z_{\rho,L}^{(2,\pm)}=\pm
 1/2$ ($z_{\sigma,L}^{(2,\pm)}=\pm 1/2$). The excited states
 $\ket{\Psi_{\nu,1}^{\pm}}$ are obtained under antiperiodic boundary
 conditions with wave number $k=\pi$ ($k=0$) for $\nu=\rho$
 ($\nu=\sigma$).  If we calculate the first excited state without
 classifying the Hilbert space by parity, the value changes
 discontinuously between $\pm1/2$ at these transition points (magenta
 line).
 On the other hand, $z_{\nu,L}^{(2)}$ changes continuously and becomes
 zero at the transition points (green line).}\label{fig:z1_tUV}
\end{figure}

\begin{figure}[t]
\begin{center}
\includegraphics[width=7.5cm]{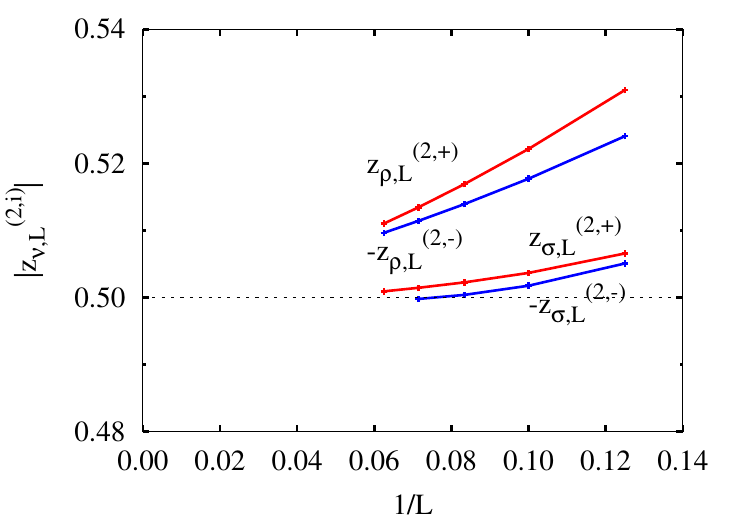}
\end{center}
\caption{System size dependence ($L=8$-$16$) of $z_{\rho,L}^{(2,\pm)}$
of the the extended Hubbard model at the BCDW-CDW boundary $V_{\rm
c}/t=1.650$ and $z_{\sigma,L}^{(2,\pm)}$ at the SDW-BCDW boundary
$V_{\rm c}/t=1.353$ at $U/t=3$ \cite{Nakamura1999,Nakamura2000}. This
shows that $z_{\nu,L}^{(2,\pm)}$ has the size dependence $O(1/L)$, and
approach to $\pm 1/2$ in the $L\to\infty$ limit.}
\label{fig:size_z1_tUV}
\end{figure}

To apply our argument to the electron system, we introduce the twist
operators for the charge and the spin sectors as\cite{Yamanaka-O-A} 
\begin{equation}
U_{\rho}\equiv U_{\uparrow}U_{\downarrow},\qquad
U_{\sigma}\equiv U_{\uparrow}U_{\downarrow}^{-1},
\end{equation}
where $U_{s}\equiv\exp[(2\pi{\rm i}/L)\sum_{j=1}^L jn_{js}]$.  Their
ground-state expectation values
$z_{\nu,L}^{(2)}\equiv\braket{\Psi_{0}|U_{\nu}|\Psi_{0}}$
($\nu=\rho,\sigma$) give the BCDW-CDW ($z_{\rho,L}^{(2)}=0$) and the
SDW-BCDW ($z_{\sigma,L}^{(2)}=0$) transition points, respectively
\cite{Nakamura-V}. In the present two-component case, the boson
representation of $U_{\nu}$ is given by (\ref{bosonized_U}) where the
phase fields for the charge ($\nu=\rho$) and the spin ($\nu=\sigma$)
sectors are replaced as $\phi\to\phi_{\nu}/\sqrt{2}$.

As shown in Fig.~\ref{fig:z1_tUV}, the expectation values of $U_{\nu}$
with respect to the excited states become $z_{\rho,L}^{(2,\pm)}=\pm 1/2$
at the BCDW-CDW transition point, and $z_{\sigma,L}^{(2,\pm)}=\pm 1/2$
at the SDW-BCDW transition point, respectively.  Their system-size
dependence at the critical points is $O(1/L)$ as shown in
Fig.~\ref{fig:size_z1_tUV}.  The excited states
$\ket{\Psi_{\nu,1}^{\pm}}$ are obtained under antiperiodic boundary
conditions $c_{i+L,s}^{\dag}=-c_{i,s}^{\dag}$ with wave number $k=\pi$
($k=0$) for the charge $\nu=\rho$ (spin $\nu=\sigma$) sector
\cite{Nakamura1999,Nakamura2000}.

In the present case, $z_{\rho,L}^{(2,\pm)}$ and $z_{\sigma,L}^{(2,\pm)}$
behave similarly to those of the $S=1$ spin chain and the $S=1/2$
$J_1$-$J_2$ spin chain, respectively, reflecting U(1) and SU(2)
symmetries of the universality class of the transitions.  The difference
of the signs of $z_{\sigma,L}^{(2,\pm)}$ and the $S=1/2$ $J_1$-$J_2$
model is due to that of the coupling constant of the nonlinear terms
$g$.


\section{Summary and discussion}\label{sec:summary}
In summary, we have discussed the expectation value of the LSM-type
twist operator $U^q$ with respect to excited states
$\ket{\Psi_{q/2}^{\pm}}$ that accompany momentum transfer $qk_{\rm
F}$. This takes the universal values $\pm 1/2$ in TL liquids, so that if
the Hilbert space of these states is not classified by discrete
symmetries, the expectation value changes discontinuously between these
two values at the phase transition points that belong to the
universality class of the U(1) or SU(2) symmetric Gaussian model.  As a
matter of fact, the behavior of $z_L^{(q,\pm)}$ is just like an
``enhanced'' version of $z_L^{(q)}$ which takes finite values with
different signs at the two gapped fixed points but becomes zero at the
transition point. However, the property of $z_L^{(q,\pm)}$ is
essentially different from that of $z_L^{(q)}$ in terms that
$z_L^{(q,\pm)}$ takes finite values on the limit of the gapless point.
This property is applicable to detect these phase transitions and
characterize the topology of the system. We have demonstrated these
properties in the $S=1/2$ $J_1$-$J_2$ spin chain, the $S=1$ Heisenberg
chain, and the extended Hubbard model.

In TL liquids, we can not define order parameters as ground-state
expectation values of operators, because the bosonized operator is
always written in normal-ordered form. In other words, this is the
consequence of an absence of long-range orders. Therefore, physical
information in a TL liquid is usually characterized by the dominant
exponents of the two-point correlation functions that show power-law
decay. Contrary to this, our result indicates that we can enhance the
order parameter $z_L^{(q)}$ and extract the physical information of TL
liquids if the average is taken in terms of appropriate excited states.

In the present argument, the universal values $z_L^{(q,\pm)}=\pm1/2$ in
TL liquids do not depend on the detailed boson representation of
$\mathcal{U}$ as long as the relation (\ref{UUU}) is satisfied.  On the
other hand, $z_L^{(q)}=0$ on the Gaussian point is explained by the
bosonized form (\ref{bosonized_U}) and the symmetry of the Gaussian
point under the transformation $\phi\to\phi+\pi/2q$ which reverses the
sign of the non-linear term of the sine-Gordon model.  In addition to
this, there is small size dependence of the $z_L^{(q)}=0$ point due to
the approximation to the linearized dispersion relation of the TL
model. Then, $\braket{:\e^{\i2q\phi}:}$ takes a finite value, and the
size dependence mainly stems from the factor
$\left(\frac{2\pi}{L}\right)^{q^2K}$ of Eq.~(\ref{bosonized_U}).
Recently, the size dependence of $z_L^{(q)}$ away from the $z_L^{(q)}=0$
point has been discussed\cite{Kobayashi-N-F-O}. For this case, effects
of the marginal operator should be taken into account in the present
case \cite{Furuya-N}.
%



\section{Acknowledgment}
M.~N. acknowledges the Visiting Researcher's Program of the Institute
for Solid State Physics, the University of Tokyo.
The authors are grateful to M.~Oshikawa for turning their attention to
the current issue and for discussions.
The authors also thank Y. Fuji, Y. Fukusumi, R. Kobayashi,
Y. O. Nakagawa, S. Nishimoto, Y. Tada, and S. Todo for fruitful
discussions.
This work is supported by JSPS KAKENHI Grant No. 17K05580.

\pagebreak

\appendix

\begin{widetext}

\section{Derivation of Eqs.(\ref{bosonized_U}) and (\ref{UUU})}
\label{sec:deriv_boson_U}

We derive Eq.~(\ref{bosonized_U}).  By rescaling the density operators
by the TL parameter and using the Campbell-Baker-Hausdorff formula
$\e^{A+B}=\e^A\e^B \e^{-\frac{1}{2}\left[A,B\right]} =\e^B\e^A
\e^{\frac{1}{2}\left[A,B\right]}$, the normal ordered representation of
Eq.~(\ref{bosonized_U0}) is calculated as follows,
\begin{align}
 \lefteqn{
\mathcal{U}(1,K)\equiv
\exp\left[
\i(2\phi(L)-N\pi-2Q)
\right]
 =\exp\left[
-\sum_{n\neq 0} \frac{\sqrt{K}}{n}
 \e^{-\alpha \frac{\pi|n|}{L}}
 \left[\tilde{\rho}_+ (n) + \tilde{\rho}_- (n)\right]
 +\i N\pi
\right]
}
\label{exponential_1}
\\
=&\underbrace{\exp\left[
-\sum_{n>0} \frac{\sqrt{K}}{n}
 \e^{-\alpha \frac{\pi|n|}{L}}
 \left[\tilde{\rho}_+ (n) - \tilde{\rho}_- (-n)\right]
 \right]}_{\equiv\exp(\i2\phi_>)}
\underbrace{
\exp\left[
-\sum_{n<0} \frac{\sqrt{K}}{n}
 \e^{-\alpha \frac{\pi|n|}{L}}
 \left[\tilde{\rho}_+ (n) - \tilde{\rho}_- (-n)\right]
\right]}_{\equiv\exp(\i2\phi_<)}\e^{\i N\pi}\nonumber\\
&\times\exp
\biggl\{
\underbrace{-\frac{K}{2}\left[
\sum_{n>0} \frac{1}{n}
 \e^{-\alpha \frac{\pi|n|}{L}}
 \left[\tilde{\rho}_+ (n) - \tilde{\rho}_- (-n)\right],
\sum_{m<0} \frac{1}{m}
 \e^{-\alpha \frac{\pi|m|}{L}}
 \left[\tilde{\rho}_+ (m) - \tilde{\rho}_- (-m)\right]
\right]}_{-\frac{K}{2}[\i2\phi_>,\i2\phi_<]\ (*)}\biggr\},
\label{sup.04}
\end{align}
where we have redefined $\tilde{\rho}_{\pm}(n)\equiv\rho_{\pm}(p)$ with
$p=\frac{2\pi}{L}n$. The marked part in Eq.~(\ref{sup.04}) becomes
\begin{align}
(*)=&\exp\left\{-\frac{K}{2}
\sum_{n>0}\sum_{m<0} \frac{1}{nm}
\e^{-\alpha \frac{\pi(|n|+|m|)}{L}}
\left[
  \left[\tilde{\rho}_+ (n) + \tilde{\rho}_- (-n)\right],
  \left[\tilde{\rho}_+ (m) + \tilde{\rho}_- (-m)\right]
\right]\right\}\nonumber\\
=&\exp\left(-\frac{K}{2}
\sum_{n>0} \frac{1}{n^2}
\e^{-\alpha \frac{2\pi n}{L}}2n\right)
=\exp\left(-K\sum_{n>0} \frac{1}{n}\e^{-\alpha \frac{2\pi n}{L}}\right)
=\exp\left(K\log(1-\e^{-\alpha \frac{2\pi}{L}})\right)\nonumber\\
\simeq&\left(\frac{2\pi\alpha}{L}\right)^{K}.
\end{align}
Therefore Eq.~(\ref{exponential_1}) becomes
\begin{align}
\mathcal{U}(1,K)&
\simeq
\left(\frac{2\pi\alpha}{L}\right)^{K}
\exp(\i2\phi_>)\exp(\i2\phi_<)\e^{\i N\pi}
\equiv\left(\frac{2\pi\alpha}{L}\right)^{K}
:\exp\left[\i(2\phi(L)-N\pi-2Q)\right]:
\nonumber\\
&\simeq\left(\frac{2\pi\alpha}{L}\right)^{K}
:\exp\left[\i 2\phi(L)\right]:.
\end{align}
Thus we get Eq.~(\ref{bosonized_U}). This satisfies Eq.~(\ref{UUU}) as
follows,
\begin{align}
\mathcal{U}(p,K)\mathcal{U}(q,K)=
&\exp(\i2p\phi_>)\exp(\i2p\phi_<)\e^{\i pN\pi}
\exp(\i2q\phi_>)\exp(\i2q\phi_<)\e^{\i qN\pi}
\left(\frac{2\pi\alpha}{L}\right)^{(p^2+q^2)K}\nonumber\\
=&\exp(\i2p\phi_>)
\exp(\i2q\phi_>)\exp(\i2p\phi_<)\exp(\i2q\phi_<)\e^{\i(p+q)N\pi}
\left(\frac{2\pi\alpha}{L}\right)^{(p^2+q^2)K}
\exp(-[\i2q\phi_>,\i2p\phi_<])\nonumber\\
=&\exp(\i2(p+q)\phi_>)\exp(\i2(p+q)\phi_<)\e^{\i(p+q)N\pi}
\left(\frac{2\pi\alpha}{L}\right)^{(p^2+q^2+2pq)K}\nonumber\\
=&\exp(\i2(p+q)\phi_>)\exp(\i2(p+q)\phi_<)\e^{\i(p+q)N\pi}
 \left(\frac{2\pi\alpha}{L}\right)^{(p+q)^2K}\nonumber\\
=&\mathcal{U}(p+q,K).
\end{align}

\section{Alternative derivation of Eq.~(\ref{correction_to_FSS})}
\label{sec:three_point}

In Refs.~\onlinecite{Cardy86,Reinicke}, Eq.~(\ref{correction_to_FSS})
has been derived based on the transfer-matrix method. Here we give an
alternative derivation of this formula using only CFT.  We consider an
expectation value of an operator $\mathcal{O}_{j}$ in terms of excited
states in the cylindrical coordinate as
\begin{align}
 z^{iji}\equiv
 {}_{\rm cyl}\braket{\mathcal{O}_i|
 \mathcal{O}_j(\sigma)|\mathcal{O}_i}_{\rm cyl},
 \label{def_z_a1a2}
\end{align}
where $\ket{\mathcal{O}_{i}}_{\rm cyl}$ is the highest weight state
corresponding to the primary operator $\mathcal{O}_i(z,\bar z)$.  We
assume that the operators are Hermitian
$\mathcal{O}_{i}^{\dag}=\mathcal{O}_{i}^{\mathstrut}$.  Then the
counterpart of the two-dimensional plain $\ket{\mathcal{O}_i}$ and its
conjugate state are defined as
\begin{subequations} 
\begin{align}
 \ket{\mathcal{O}_i}&\equiv\lim_{z,\bar z\to 0}
 z^{-\Delta_i} \bar z^{-\bar\Delta_i}
 \mathcal{O}_{i}(z,\bar z)\ket{0},
 \label{def_ket_V_a}\\
 \bra{\mathcal{O}_i}&\equiv\lim_{z,\bar z \to  0}
 z^{-\Delta_i} \bar z^{-\bar\Delta_i}
 \bra{0} \mathcal{O}_{i}(1/z,1/\bar z),
 \label{def_bra_V_a}
\end{align}
\end{subequations}
where $(\Delta_i,\bar\Delta_i)$ is the conformal dimension of
$\mathcal{O}_i$. The above definitions satisfy the normalization
condition $\braket{\mathcal{O}_i|\mathcal{O}_i}=1$.  We now define
$\ket{\mathcal{O}_i}$ and $\bra{\mathcal{O}_i}$ on a cylinder with
length $L$ as follows:
\begin{subequations} 
\begin{align}
 \ket{\mathcal{O}_i}_{\rm cyl}
 &\equiv\lim_{w,\bar w \to -\infty}
 \biggl(\frac{L}{2\pi a z}\biggr)^{\Delta_i}
 \biggl(\frac{L}{2\pi a\bar z}\biggr)^{\bar\Delta_i}
  \mathcal{O}_{i}(w,\bar w) \ket{0}, \\
 {}_{\rm cyl}\bra{\mathcal{O}_i}
 &\equiv\lim_{w, \bar w \to -\infty}\lim_{w'\to -w} 
 \lim_{\bar w' \to -\bar w}
 \biggl(\frac{L}{2\pi a z}\biggr)^{\Delta_i}
 \biggl(\frac{L}{2\pi a\bar z}\biggr)^{\bar\Delta_i} \bra{0}
 \mathcal{O}_{i}(w', \bar w'),
\end{align}
\end{subequations}
where $a$ is the lattice constant, and $w,\bar w$ and $z,\bar z$ are
related by the conformal transformation,
\begin{align}
 w = \frac{L}{2\pi a}\ln z, \qquad \bar w = \frac{L}{2\pi a}\ln \bar z.
\end{align}
The
normalization between $\ket{\mathcal{O}_i}_{\rm cyl}$ and ${}_{\rm
cyl}\bra{\mathcal{O}_i}$ is confirmed as follows:
\begin{align}
 {}_{\rm cyl}\braket{\mathcal{O}_i |\mathcal{O}_i}_{\rm cyl}
 &= \lim_{w, \bar w \to -\infty}\lim_{w'\to -w}\lim_{\bar w'\to -\bar w}
 \biggl(\frac{L}{2\pi a z}\biggr)^{2\Delta_i}
 \biggl(\frac{L}{2\pi a \bar z} \biggr)^{2\bar\Delta_i}
 \braket{0|\mathcal{O}_{i}(w', \bar w') \mathcal{O}_i(w,\bar w)|0}
 \notag \\
 &= \lim_{z,\bar z\to 0}\biggl(\frac{L}{2\pi a
 z}\biggr)^{2\Delta_i}
 \biggl(\frac{L}{2\pi a\bar z} \biggr)^{2\bar\Delta_i}
 \notag \\
 &\qquad \times \lim_{z'\to 1/z}\lim_{\bar z'\to 1/\bar z}
 \biggl(\frac{L}{2\pi a z'}\biggr)^{-\Delta_i}
 \biggl(\frac{L}{2\pi a \bar z'}\biggr)^{-\bar\Delta_i}
 \biggl(\frac{L}{2\pi a z}\biggr)^{-\Delta_i}
 \biggl(\frac{L}{2\pi a \bar z}\biggr)^{-\bar\Delta_i}
 \braket{0|\mathcal{O}_{i}(z', \bar z')\mathcal{O}_i(z,\bar z)|0}
 \notag \\
 &= \lim_{z,\bar z\to 0} \biggl( \frac 1z\biggr)^{2\Delta_i}
 \biggl( \frac 1{\bar z}\biggr)^{2\bar\Delta_i}
 \frac{1}{(1/z)^{2\Delta_i}(1/\bar z)^{2\bar\Delta_i}}
 \notag \\
 &= 1.
\end{align}
The expectation value of
$\mathcal{O}_{j}(\sigma)=\mathcal{O}_{j}(w,\bar{w})$ with
$w=\tau+\i\sigma$, $\bar{w}=\tau-\i\sigma$ in terms of
$\ket{\mathcal{O}_i}_{\rm cyl}$ is calculated as follows,
\begin{align}
 z^{iji}
 &= {}_{\rm cyl}\braket{\mathcal{O}_{i}|
 \mathcal{O}_{j}(w,\bar{w})|\mathcal{O}_{i}}_{\rm cyl}
 \notag \\
 &=
 \lim_{w'', \bar w''\to -\infty} \lim_{w'\to -w''}
 \lim_{\bar w'\to -\bar w'}
 \biggl(\frac{L}{2\pi a z''}\biggr)^{2\Delta_i}
 \biggl(\frac{L}{2\pi a\bar z''}\biggr)^{2\bar\Delta_i}
 \braket{0|\mathcal{O}_{i}(w', \bar w')
 \mathcal{O}_{j}(w,\bar{w})
 \mathcal{O}_{i}(w'',\bar w'')|0}
 \notag \\
 &= \lim_{z'',\bar z''\to 0} \lim_{z'\to 1/z''}\lim_{\bar z'\to 1/\bar z''}
 \biggl(\frac{L}{2\pi az''}\biggr)^{2\Delta_i}
 \biggl(\frac{L}{2\pi a\bar z''}\biggr)^{2\bar\Delta_i}
 \notag \\
 &\qquad \times
 \biggl(\frac{L}{2\pi a z'}\biggr)^{-\Delta_i}
 \biggl(\frac{L}{2\pi a \bar z'}\biggr)^{-\bar\Delta_i}
 \biggl( \frac{L}{2\pi az}\biggr)^{-\Delta_{j}}
 \biggl(\frac{L}{2\pi a\bar{z}}\biggr)^{-\bar\Delta_{j}}
 \biggl(\frac{L}{2\pi a z''}\biggr)^{-\Delta_i}
 \biggl(\frac{L}{2\pi a \bar z''}\biggr)^{-\bar\Delta_i}
 \notag \\
 &\qquad \times \braket{0|\mathcal{O}_{i}(z', \bar z')
 \mathcal{O}_{j}(z,\bar{z}) \mathcal{O}_{i}(z'',\bar z'')|0}
 \notag \\
 &= \lim_{z'',\bar z''\to 0}
 \biggl(\frac{2\pi a}{L}\biggr)^{x_j}
 \frac{z^{\Delta_j}\bar z^{\bar\Delta_j}}
 {z''^{2\Delta_i}\bar z''^{2\bar\Delta_i}}
 \braket{0|\mathcal{O}_{i}(1/z'', 1/\bar z'')
 \mathcal{O}_{j}(z,\bar{z}) \mathcal{O}_{i}(z'',\bar z'')|0}
 \label{three_point_func}\\
 &=\biggl(\frac{2\pi a}{L}\biggr)^{x_j}C_{iji},
\end{align}
where $x_j=\Delta_j+\bar\Delta_j$ is the scaling dimension of
$\mathcal{O}_{j}$, and $C_{iji}$ is the OPE coefficient of the three
point function in Eq.~(\ref{three_point_func}).  Thus we get
Eq.~(\ref{correction_to_FSS}).

\section{Operator product expansion coefficients}
\label{sec:OPE}

We calculate OPE coefficients involving the following operators,
\begin{equation}
  {\cal O}_1(\sigma)=:\cos[q\phi(\sigma)]:,\qquad
  {\cal O}_2(\sigma)=:\sin[q\phi(\sigma)]:,\qquad
  {\cal O}_3(\sigma)=:\cos[2q\phi(\sigma)]:.
\end{equation}
In a spin-$1/2$ chain, ${\cal O}_1$ and ${\cal O}_2$ correspond to the
singlet state and the triplet state with $S^z=0$, respectively.  ${\cal
O}_3$ appears in the umklapp scattering term. The phase field is given
by the holomorphic and the antiholomorphic parts as,
\begin{equation}
 \phi(z,\bar{z}) = \frac{\sqrt{K}}{2}
  [\varphi(z)+ \bar{\varphi}(\bar{z})].
 \label{varphi}
\end{equation}
The vertex operators satisfy the following OPE rule for $z\simeq z'$:
\begin{equation}
 :\e^{\i \alpha \varphi (z)}::\e^{\i \beta \varphi (z')}:
  \simeq(z-z')^{\alpha \beta }
  :\e^{\i(\alpha+\beta)\varphi(z')}:.
  \label{OPE_V-V}
\end{equation}
Then the OPE of $\mathcal{O}_2$, $\mathcal{O}_3$ is given by their most
divergent terms as
\begin{align}
 \mathcal{O}_2(z, \bar{z}) \mathcal{O}_3(z', \bar{z}')
 \simeq&
 \lefteqn{\frac{1}{4\i}\left(
  :\e^{\i q\sqrt{K/4}\varphi(z)}:
  :\e^{\i q\sqrt{K/4}\bar{\varphi}(\bar{z})}:
  :\e^{-\i q\sqrt{K}\varphi(z')}:
  :\e^{-\i q\sqrt{K}\bar{\varphi}(\bar{z}')}:-{\rm H.c.}\right)}\nonumber\\
 \simeq&\frac{1}{4\i}\frac{1}{(z-z')^{q^2K/2}(\bar{z}-\bar{z}')^{q^2K/2}}
  \left(:\e^{-\i q\sqrt{K/4}\varphi(z')}:
   :\e^{-\i q\sqrt{K/4}\bar{\varphi}(\bar{z}')}:
 -{\rm H.c.}\right)\nonumber\\
 =&\frac{-1/2}{(z-z')^{q^2K/2}(\bar{z}-\bar{z}')^{q^2K/2}}
 \mathcal{O}_2 (z',\bar{z}').
\end{align}
Thus, we obtain $C_{232}=-1/2$. Similarly, we obtain $C_{131}=1/2$ as
\begin{equation}
 \label{ope.20}
  \mathcal{O}_1(z,\bar{z}) \mathcal{O}_3(z',\bar{z}')
  \simeq\frac{1/2}{(z-z')^{q^2K/2} (\bar{z}-\bar{z}')^{q^2K/2}}
  \mathcal{O}_1(z',\bar{z}').
\end{equation}
The process to obtain these universal values $C_{131}=1/2$ and
$C_{232}=-1/2$ is quite similar to that of Eq.~(\ref{z03}).
\end{widetext}
\end{document}